\documentclass[twocolumn,showpacs,preprintnumbers,amsmath,amssymb]{revtex4}
\usepackage{epsfig}
\preprint{PRL-214/2003}
\begin{document}

\textbf{Lucarelli et al. reply to Tajima et al.}

In their Comment on our paper,\cite{Lucarelli} Tajima \textit{et al.}
\cite{Tajima} argue that our infrared results on
nine single crystals of La$_{2-x}$Sr$_x$CuO$_4$ (LSCO) are not valid because:

i) Ref. [1] reports three peaks, at 30 (for $x$ = 0.12), 250, and
500 cm$^{-1}$; ii) the $x$ = 0.05 sample shows
a dip at 470 cm$^{-1}$ due to a tranverse optical phonon of the $c$-axis at 500 cm$^{-1}$; therefore, the sample is not a
single crystal or it is miscut; iii) the same dip is observed
more or less in all samples, except for those with $x$ = 0.0  and
0.26; therefore, all samples are bad crystals, are
miscut, or the polarizer was not effective; iv) as most samples
contain the $c$-axis, also the peak at 30 cm$^{-1}$ is a spurious
feature; v) \textit{the} previous observations on the same
system do not show the peaks we report in Ref. [1].

Details on the samples and experiments will be given in a forthcoming paper.\cite{prb} Here, we reply to each of the above points.

i) In \textit{all} the superconducting crystals investigated in Ref. [1], strong peaks are observed \textit{below $\sim$ 150} cm$^{-1}$, which depend on temperature and doping consistently with a charge stripe
scenario (Figs. 2 and 3 of Ref. [1]). The strong conductivity peak at 250 
cm$^{-1}$ and the weak peak at 500 cm$^{-1}$ are discussed only for the semiconducting 0.05 sample, in connection with Thomas \textit{et al.}
who observed very similar features in a flux-grown La$_2$SrCuO$_{4+y}$ crystal where the surface is intrinsically $a-b$.\cite{Thomas}

ii) The $x$ = 0.05 sample arrived already cut. We verified that it is an excellent single crystal by X-ray measurements at the Walther-Meissner Institut of Garching (Germany).\cite{prb} 

iii) This is a crucial point, as it questions the peaks below 150 cm$^{-1}$. We are surprised by this comment. Fig. 1 shows the reflectivity $R({\omega})$ for our $x$ = 0.12 sample and for an $x$ = 0.13 LSCO crystal
from a paper co-authored by one (D. N. B.) of the
authors of the Comment.\cite{Startseva1} Both
samples show a dip at 470 cm$^{-1}$ for electric field orthogonal to the $c$ axis. In Ref. \onlinecite{Startseva1}, the 0.13 sample is described as
follows: "The miscut angle between the polished surface and the
$c$ axis was checked by a high precision triple axis X-ray
diffractometer and was determined to be less than 0.8$^0$".
Therefore, in no way the presence of the dip at 470 cm$^{-1}$ can
be used as evidence for a miscut of the crystal. This feature has been observed, indeed, in flux-grown La$_2$CuO$_{4+y}$,\cite{Tanner} in well-cut (error less than 1$^0$) LSCO\cite{Shimada} and, with minor changes, in many other cuprates where it has been explained in a non-trivial way.\cite{Reedyk}

iv) In Fig. 1 one sees also that the low-frequency side of the $a-b$ plane $R({\omega})$, where the strong peaks appear, is not affected by the $c$-axis reflectivity. The same holds for the other samples.\cite{prb} Incidentally, in the $x$ = 0.15 spectrum,[1] strong far-infrared peaks spring up at low $T$ while the dip at 470 cm$^{-1}$ is very weak at any $T$.

v) The authors of Ref. [2] support their argument by citing three
papers. Two of them report data on thin films, known to be affected by the subtraction of the
substrate contribution, the third one reports measurements at grazing
incidence. When measuring at quasi-normal incidence single LSCO crystals, anomalous peaks are observed instead, similar to those in Ref. [1]: see, therein, Refs. [16],[21],[22], and a recent preprint as well.\cite{Kim} Concerning the paper of Dumm \textit{et al.},\cite{Dumm} the remark in Ref. [2] does not report faithfully the note we added to Ref. [1].

In conclusion, the high quality and good orientation of the crystals measured in Ref. [1] is not questioned by the presence of a dip in $R({\omega})$ at 470 cm$^{-1}$. The latter feature appears both in well-cut crystals grown by the floating-zone method and in flux-grown samples. Our case on a charge-stripe scenario for LSCO is supported by a sound trend of the optical conductivity throughout its phase diagram and fully agrees with the results of Raman\cite{Venturini} and neutron\cite{Fujita} spectroscopy on samples from the same laboratories.

A. Lucarelli, S. Lupi, M. Ortolani, P. Calvani, P. Maselli, and M. Capizzi

"Coherentia"-INFM and Dipartimento di Fisica, Universit\`a di Roma La Sapienza, Roma, Italy.

\begin{figure}
{\hbox{\epsfig{figure=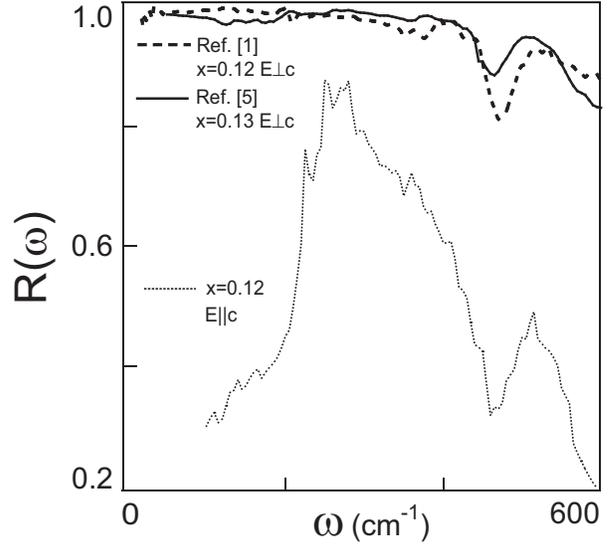,width=8cm}}}
\caption{Reflectivity for field orthogonal to the $c$ axis for our sample with $x$ = 0.12 at 30 K and for a sample with $x$ = 0.13 at 10 K from Ref. [5]. In the latter crystal the miscut was less than 0.8$^0$. A test of $R({\omega})$ along $c$ at 295 K for the same $x$ = 0.12 sample is also shown.}
\label{refl}
\end{figure}

\end{document}